\documentclass[12pt]{article}

\usepackage[compact]{titlesec}
\usepackage{mathtools,bbm}
\usepackage{graphicx,fullpage,setspace,amsmath,amsthm,textcomp,amsfonts,amssymb,verbatim,url,authblk,xcolor,lipsum}
\usepackage{float,rotating,longtable,epsfig,color,multirow,lscape,caption,soul}
\usepackage{bm}
\usepackage[comma, authoryear]{natbib}

\setstcolor{red}

\title{On equivalence of the LKJ distribution and the restricted Wishart distribution}

\author[1$\dagger$]{Zhenxun Wang}
\author[2]{Yunan Wu}
\author[1]{Haitao Chu}

\affil[1]{Division of Biostatistics, School of Public Health,
	The University of Minnesota, Minneapolis, MN 55455}
\affil[2]{School of Statistics, The University of Minnesota, Minneapolis, MN 55455}
\affil[$\dagger$]{Corresponding author:  wang6795@umn.edu}
\date{}

\begin{document}
\thispagestyle{empty}
\setcounter{page}{0}
	
\maketitle
	
\clearpage
\setcounter{page}{0}
\begin{abstract}
In this short paper, we want to show the Restricted Wishart distribution is equivalent to the LKJ distribution, which is one way to specify a uniform distribution from the space of positive definite correlation matrices [\cite{Lewandowski2009}]. Based on this theorem, we propose a new method to generate random correlation matrices from the LKJ distribution. This new method is faster than the original onion method for generating random matrices, especially in the low dimension ($T<120$) situation.


\end{abstract}
\clearpage

\section{Introduction}
\label{intro}
The choice of prior distribution for the correlation or covariance matrix is crucial but difficult in Bayesian Analysis. It is challenging because the number of parameters in the covariance matrix increases rapidly as the dimension of the matrix increases and these parameters are constrained by the complicated condition that the matrix should be non-negative definite (\citealp{Barnard2000}). It's also hard because we have little intuition about how the entries in the matrix should be correlated a priori. The most traditional choice is the inverse Wishart distribution, the conjugate prior for the covariance matrix of a multivariate normal distribution. However, such a prior has its own problems, for example, the marginal distribution of the variances has low density in a region near zero (\citealp{Gelman2006}). An increasing popular alternative is the separation strategy proposed by \cite{Barnard2000}, which separates the variances and the correlation matrix and considers them independently. To sample the correlation matrix uniformly from the space of positive definite correlation matrices, Lewandowski et al.(\citeyear{Lewandowski2009}) proposed the LKJ distribution, which is based on the transformation of partial correlations to the correlation matrix. The LKJ distribution is now widely used and is the default prior for a correlation matrix in the STAN software (\citealp{Carpenter2017}). In this paper, we will apply the separation strategy to the Wishart distribution instead of to the inverse-Wishart distribution in Barnard et al (\citeyear{Barnard2000}). We call the new distribution the \textbf{Restricted Wishart Distribution}. In addition, we show that the restricted Wishart distribution is equivalent to the LKJ distribution.

The rest of paper is organized as follows. Section 2 gives a brief introduction to the inverse Wishart distribution and the separation strategy. Then Section 3 applies this strategy to the Wishart distribution and shows the equivalence of the restricted Wishart distribution and the LKJ distribution. Finally, we propose a new method to generate random matrices from the LKJ distribution and compare the speed of the different methods.
\section{Inverse Wishart prior and the separation strategy} \label{Inverse Wishart prior}
The inverse Wishart (IW) prior is the conjugate prior for the covariance matrix in a multivariate normal model. Specifically, if $T\times 1$ vectors $\bm{\nu_{1}}, \bm{\nu_{2}}, \ldots, \bm{\nu_{K}} \stackrel{iid}{\sim} MVN(\boldsymbol{0},\boldsymbol{\Sigma})$ and we have a $T\times K$ matrix
$\bm{\nu} = (\bm{\nu_{1}}, \bm{\nu_{2}}, \ldots, \bm{\nu_{K}})$, then the $T$-dimensional positive definite matrix $\boldsymbol{S} = \bm{\nu} \bm{\nu}^{\prime}$ ($\bm{\nu}^{\prime}$ is the transpose of $\bm{\nu}$) follows the Wishart distribution with degree of freedom $K>T-1$ and positive definite scale matrix $\boldsymbol{\Sigma}$:
\begin{align}\label{dist:W}
p(\boldsymbol{S})=W_{T}(K,\boldsymbol{\Sigma}) \equiv \{2^{\frac{1}{2}KT}\Gamma_{T}(\frac{1}{2}K)\}^{-1}|\boldsymbol{\Sigma}|^{-\frac{1}{2}K}|\boldsymbol{S}|^{\frac{1}{2}(K-T-1)}\exp(-\frac{1}{2}tr(\boldsymbol{\Sigma}^{-1}\boldsymbol{S}))
\end{align}
where $|\bullet|$ is the determinant, $tr$ is the trace, and $\Gamma_{T}$ is the multivariate (T-variate here) gamma function. Then the conjugate prior for $\boldsymbol{\Sigma}$ will be inverse-Wishart Distribution with degree of freedom $m>T-1$ and positive definite scale matrix $\boldsymbol{\Psi}$:
\begin{align}\label{dist:IW}
\pi(\boldsymbol{\Sigma})=IW_{T}(m, \boldsymbol{\Psi}) \propto |\boldsymbol{\Psi}|^{-\frac{1}{2}m}|\boldsymbol{\Sigma}|^{-\frac{1}{2}(m+T+1)}\exp(-\frac{1}{2}tr(\boldsymbol{\Sigma}^{-1}\boldsymbol{\Psi}))
\end{align}
The inverse of $\boldsymbol{\Sigma}$ has a Wishart distribution $\pi(\boldsymbol{\Sigma}^{-1}) \propto W_{T}(m,\boldsymbol{\Psi}^{-1})$. According to the separation strategy, we can further decompose $\boldsymbol{\Sigma}$
as $\boldsymbol{\Delta} \boldsymbol{P} \boldsymbol{\Delta}$, where $\boldsymbol{\Delta}$ is a diagonal matrix with standard deviation $\delta_i=\sqrt{\sigma_{ii}}$ as its $i^{th}$ diagonal
element and $\boldsymbol{P}$ is a correlation matrix with diagonal elements 1 and off-diagonal elements $\rho_{ij}$.
In addition, the marginal distribution of a principal sub-matrix of an inverse-Wishart random variable is still distributed as inverse-Wishart (\citealp{Barnard2000}). That is, for any $T_1 \times T_1$ principal sub-matrix $\boldsymbol{\Sigma}_{1}$ of $\boldsymbol{\Sigma}$
\begin{align} 
\pi(\boldsymbol{\Sigma}_{1}) \propto IW_{T_1}(m-T+T_1,\boldsymbol{\Psi}_1)
\end{align}
where $\boldsymbol{\Psi}_1$ is the $T_1 \times T_1$ principal sub-matrix of $\boldsymbol{\Psi}$, the same sub-matrix as $\boldsymbol{\Sigma}_{1}$ is of $\boldsymbol{\Sigma}$. Simply letting $T_1=1$, we find the marginal distribution of $\sigma_{ii}=\delta_{i}^{2}$ is exactly the inverse-gamma distribution
\begin{align} \label{dist:IG}
\pi(\sigma_{ii}) = IG(\frac{m-T+1}{2},\frac{\psi_{ii}}{2}), \ \ i = 1, \ldots, T
\end{align}
where $\psi_{ii}$ is the $ii^{th}$ entry of $\boldsymbol{\Psi}$.
When $\boldsymbol{\Psi}$ is a diagonal matrix $diag(\psi_{11}, \ldots, \psi_{TT})$, we can derive the marginal distribution of $\boldsymbol{P}$, following the approach used in \cite{Barnard2000} to derive the marginal distribution of $\boldsymbol{P}$ when $\boldsymbol{\Psi}$ is the identity matrix. We first calculate the Jacobian matrix of the transformation $\boldsymbol{\Sigma} \rightarrow (\sigma_{11}, \ldots, \sigma_{TT},\boldsymbol{P})$

\begin{align} \label{J_1}
\boldsymbol{J}_1 =
\boldsymbol{J}(\sigma_{11}=\sigma_{11}, \ldots, \sigma_{TT}=\sigma_{TT}, \rho_{ij}=\frac{\sigma_{ij}}{\sqrt{\sigma_{ii}\sigma_{jj}}}) =
\left[
\begin{array}{c|c}
\boldsymbol{I}_{T} & 0 \\
\hline
* & \boldsymbol{C}
\end{array}
\right]
\end{align}
where $\boldsymbol{I}_{T}$ is the $T$-dimensional identity matrix, $\boldsymbol{C}$ is the $T(T-1)/2$ dimensional diagonal matrix with entries $1/\sqrt{\sigma_{ii}\sigma_{jj}}$ ($i \neq j$). Since $J_1$ is a lower triangular matrix, its determinant $|\boldsymbol{J}_1|$ equals $\prod_{i=1}^{T}\sigma_{ii}^{-(T-1)/2}$ and sub-matrix in (\ref{J_1}) given as $*$ need not be derived. We also can calculate the Jacobian of the transformation $(\sigma_{11}, \ldots, \sigma_{TT}) \rightarrow (\delta_{1}, \ldots, \delta_{T})$. Its determinant $|\boldsymbol{J}_2|$ is $2^{-T}\prod_{i=1}^{T}\sigma_{ii}^{-1/2}$.
Hence, the joint distribution of $\boldsymbol{P}$ and $(\delta_{1}, \ldots, \delta_{T})$ is
\begin{align}
\pi(\boldsymbol{P},\delta_{1}, \ldots, \delta_{T}) &\propto \pi(\boldsymbol{\Sigma})|\boldsymbol{J}_1|^{-1}|\boldsymbol{J}_2|^{-1} \nonumber \\
&\propto |\boldsymbol{P}|^{-\frac{1}{2}(m+T+1)}\prod_{i=1}^{T}(\delta_{i}^{-(m+1)}\exp(-\frac{\rho^{ii}\psi_{ii}}{2\delta_{i}^{2}}))
\end{align}
where $\rho^{ii}$ is the $i^{th}$ diagonal element of $\boldsymbol{P}^{-1}$. Clearly, there is some dependence between the standard deviations $\delta_{i}$ and the correlation matrix $\boldsymbol{P}$. If we let $\epsilon_{i}=\rho^{ii}\psi_{ii}/(2\delta_{i}^{2})$, then the marginal distribution of $\boldsymbol{P}$ is
\begin{align}
\pi(\boldsymbol{P}) &\propto |\boldsymbol{P}|^{-\frac{1}{2}(m+T+1)}\prod_{i=1}^{T}\int_{0}^{\infty}\delta_{i}^{-(m+1)}\exp(-\frac{\rho^{ii}\psi_{ii}}{2\delta_{i}^{2}}) d\delta_{i} \nonumber \\
&\propto |\boldsymbol{P}|^{-\frac{1}{2}(m+T+1)}(\prod_{i=1}^{T}\rho^{ii}\psi_{ii})^{-\frac{m}{2}}\prod_{i=1}^{T}\int_{0}^{\infty}\epsilon_{i}^{(m-2)/2}\exp(-\epsilon_{i}) d\epsilon_{i}
\end{align}
Hence, given the fact $\rho^{ii} \equiv |\boldsymbol{P}_{ii}|/|\boldsymbol{P}|$,
\begin{align}
\pi(\boldsymbol{P}) \propto |\boldsymbol{P}|^{\frac{1}{2}(m-1)(T-1)-1}(\prod_{i=1}^{T}|\boldsymbol{P}_{ii}|)^{-\frac{m}{2}}
\end{align}
with $\boldsymbol{P}_{ii}$ being the $i^{th}$ principal sub-matrix of $\boldsymbol{P}$. We call this the \textbf{restricted Inverse-Wishart distribution}: $\boldsymbol{P} \sim RIW_{T}(m)$. Using the same approach that was used to derive equation (\ref{dist:IG}), simply set $T_1=2$ we get the marginal distribution of $\rho_{ij}$
\begin{align}
\pi(\rho_{ij}) \propto (1-\rho_{ij}^{2})^{\frac{m-T-1}{2}}, \ \ -1 \leq \rho_{ij} \leq 1
\end{align}
which is exactly the beta distribution $Beta(\frac{m-T+1}{2},\frac{m-T+1}{2})$ on $[-1,1]$, and will be uniform when $m = T+1$.

\section{Applying the strategy to the Wishart prior}
\cite{Chung2015} also recommended the Wishart prior on covariance matrix $\boldsymbol{\Sigma}$,
\begin{align}
\pi(\boldsymbol{\Sigma})=W_{T}(m,\boldsymbol{\Psi}) \equiv \{2^{\frac{mT}{2}}\Gamma_{T}(\frac{m}{2})\}^{-1}|\boldsymbol{\Psi}|^{-\frac{m}{2}}|\boldsymbol{\Sigma}|^{\frac{m-T-1}{2}}\exp(-\frac{1}{2}tr(\boldsymbol{\Psi}^{-1}\boldsymbol{\Sigma})), \ m>T-1
\end{align}
with $m=T+2$ and $\boldsymbol{\Psi}$ being the identity matrix multiplied by a large value (e.g. $a=10000$). Here we want to show how this Wishart prior is related to the jointly uniform prior for correlation matrix $\boldsymbol{P}$ (\citealp{Barnard2000}) and the LKJ distribution proposed by (\citealp{Lewandowski2009}). Actually, we want to show that given $\boldsymbol{\Psi} = diag(\psi_{11}, \ldots, \psi_{TT})$, the correlation matrix $\boldsymbol{P}$ and the variances $\sigma_{11}, \ldots, \sigma_{TT}$ of $\boldsymbol{\Sigma}$ are independently distributed. We follow an approach similar to the one in Section \ref{Inverse Wishart prior}. Given the determinant of the Jacobian $|\boldsymbol{J}(\boldsymbol{\Sigma} \rightarrow \sigma_{11}, \ldots, \sigma_{TT},\boldsymbol{P})| = |\boldsymbol{J}_1|=\prod_{i=1}^{T}\sigma_{ii}^{-(T-1)/2}$ (equation (\ref{J_1})), the joint distribution of $\boldsymbol{P}$ and $(\sigma_{11}, \ldots, \sigma_{TT})$ is
\begin{align}
\pi(\boldsymbol{P},\sigma_{11}, \ldots, \sigma_{TT}) & =  \pi(\boldsymbol{\Sigma})|\boldsymbol{J}_1|^{-1} \nonumber \\
& = 
\frac{\Gamma^{T}(\frac{m}{2})|\boldsymbol{P}|^{\frac{m-T-1}{2}}}{\Gamma_{T}(\frac{m}{2})} \prod_{i=1}^{T}\{ \frac{\sigma_{ii}^{\frac{m}{2}-1}\exp(-\frac{\sigma_{ii}}{\psi_{ii}})}{2^{\frac{m}{2}} \psi_{ii}^{\frac{m}{2}}\Gamma(\frac{m}{2})} \}
\end{align}
Clearly, $\frac{\sigma_{ii}}{\psi_{ii}}, i=1, \ldots, T$ are independently distributed as chi-square with m degrees of freedom. Also, the density of $\boldsymbol{P}$ is
\begin{align} \label{RW}
\pi(\boldsymbol{P}) = \frac{\Gamma^{T}(\frac{m}{2})}{\Gamma_{T}(\frac{m}{2})}|\boldsymbol{P}|^{\frac{m-T-1}{2}}
\end{align}
We call this the \textbf{restricted Wishart distribution}: $\boldsymbol{P} \sim RW_{T}(m)$. Apparently, the Wishart prior with $m=T+2$ and $\boldsymbol{\Psi}=a\boldsymbol{I}$ is equivalent to the separation strategy with a $RW_{T}(T+2)$ prior on the correlation matrix and a vague prior on the variances $a\times\chi^2_{T+2}$ (with large $a$). If $m=T+1$, then $\pi(\boldsymbol{P}) \propto 1$, which is the jointly uniform prior on a compact subspace of the $T(T-1)/2$ dimensional hypercube $[-1,1]^{T(T-1)/2}$ (\citealp{Barnard2000}). Instead of the method discussed here, Barnard et al. (\citeyear{Barnard2000}) used the greedy Gibbs sampler to make draws from this prior. We also can get the marginal distribution of entry $\rho_{ij}$ using the following theorem. For any $T_1 \times T_1$ principal sub-matrix $\boldsymbol{\Sigma}_{1}$ of $\boldsymbol{\Sigma}$
\begin{align}
\pi(\boldsymbol{\Sigma}_{1}) \propto W_{T_1}(m,\boldsymbol{\Psi}_1)
\end{align}
with $\boldsymbol{\Psi}_1$ being the $T_1 \times T_1$ principal sub-matrix of $\boldsymbol{\Psi}$ \cite[p.~256]{Eaton1983}. In particular, let $T_1$=2, we obtain the marginal distribution of $\rho_{ij}$
\begin{align}
\pi(\rho_{ij}) \propto (1-\rho_{ij}^{2})^{\frac{m-3}{2}}, \ \ -1 \leq \rho_{ij} \leq 1
\end{align}
which is exactly a $Beta(\frac{m-1}{2},\frac{m-1}{2})$ on $[-1,1]$, and will be $Beta(\frac{T}{2},\frac{T}{2})$ when $\pi(\boldsymbol{P}) \propto 1$.

Next we want to prove the equivalence of the restricted Wishart distribution and the LKJ distribution. The explicit density function of the LKJ distribution is following (\citealp{Lewandowski2009})
\begin{align} \label{LKJ}
\pi(\boldsymbol{P}) &= c_d^{-1}|\boldsymbol{P}|^{\alpha_{d-1}-1} \\
c_d &= 2^{\sum_{k=1}^{d-1}(2\alpha_{d-1}-2+d-k)(d-k)} \times \prod_{k=1}^{d-1}[B(\alpha_{d-1}+\frac{d-1-k}{2},\alpha_{d-1}+\frac{d-1-k}{2})]^{d-k} \nonumber
\end{align} 
where $B(\bullet)$ is the beta function, $d=T$ and $\alpha_{d-1}=(m-T+1)/2$ by changing notation. Then, equation (\ref{RW}) equals to equation (\ref{LKJ}) if and only if the constants are same
\begin{align} \label{constant}
\frac{\Gamma^{T}(\frac{m}{2})}{\Gamma_{T}(\frac{m}{2})} \times c_d = 1 .
\end{align}
Define the left hand side of above as $f(T,m)$. Then, given $\Gamma_{T}(\frac{m}{2}) = \pi^{T(T-1)/4}\prod_{k=1}^{T}\Gamma(\frac{m}{2}+\frac{1-k}{2})$ \cite[p.~483]{James1964}, we can simplify $f(T,m)$ as 
\begin{align} \label{f(t,m)}
f(T,m) &= \frac{\Gamma^{T}(\frac{m}{2})}{\Gamma_{T}(\frac{m}{2})} \prod_{k=1}^{T-1}[B(\frac{m-k}{2},\frac{m-k}{2})]^{T-k} \times
2^{\sum_{k=1}^{T-1}(m-k-1)(T-k)} \nonumber \\
&= \frac{\Gamma^{T}(\frac{m}{2})}{\pi^{T(T-1)/4}\Gamma(\frac{m}{2})\prod_{k=1}^{T-1}\Gamma(\frac{m-k}{2})} 
\prod_{k=1}^{T-1}[\frac{\Gamma^{2}(\frac{m-k}{2})}{\Gamma(m-k)}]^{T-k} \times
2^{\sum_{k=1}^{T-1}(m-k-1)(T-k)} \nonumber \\
&= \prod_{k=1}^{T-1} [2^{(m-k-1)(T-k)} \times 
\frac{\Gamma(\frac{m}{2})}{\pi^{T/4}}
\frac{\Gamma^{2T-2k-1}(\frac{m-k}{2})}{\Gamma^{T-k}(m-k)}]
\end{align}
\begin{proof}[\unskip\nopunct]
We prove that (\ref{f(t,m)}) equals 1 using mathematical induction. Start with $T=2$, then for any $m>1$, $f(2,m)$ reduces to
\begin{align} \label{dup_gamma} 2^{m-2}\pi^{-\frac{1}{2}}\frac{\Gamma(\frac{m}{2})\Gamma(\frac{m-1}{2})}{\Gamma(m-1)}
\equiv 1
\end{align}
which is known as the duplication formula of the gamma function \cite[p.~256]{1965}.
Assume $f(T,m)=1$ holds for $T=t$. It must then be shown that $f(t+1,m)=1$, $\forall m>T-1$, where
\begin{align}
f(t+1,m) &= \prod_{k=1}^{t} [2^{(m-k-1)(t+1-k)} \times 
\frac{\Gamma(\frac{m}{2})}{\pi^{(t+1)/4}}
\frac{\Gamma^{2t-2k+1}(\frac{m-k}{2})}{\Gamma^{t+1-k}(m-k)}]  \nonumber \\
&= [2^{m-t-1}\frac{\Gamma(\frac{m}{2})}{\pi^{(t+1)/4}}
\frac{\Gamma(\frac{m-t}{2})}{\Gamma(m-t)}] 
\times 
\prod_{k=1}^{t-1}[2^{(m-k-1)(t+1-k)} \times 
\frac{\Gamma(\frac{m}{2})}{\pi^{(t+1)/4}}
\frac{\Gamma^{2t-2k+1}(\frac{m-k}{2})}{\Gamma^{t+1-k}(m-k)}]  \nonumber \\
&= [2^{m-t-1}\frac{\Gamma(\frac{m}{2})}{\pi^{(t+1)/4}}
\frac{\Gamma(\frac{m-t}{2})}{\Gamma(m-t)}] 
\times f(t,m) \times 
\prod_{k=1}^{t-1} [2^{m-k-1} 
\frac{1}{\pi^{1/4}}
\frac{\Gamma^{2}(\frac{m-k}{2})}{\Gamma(m-k)}]
\end{align}
Given the identity $\Gamma(m-k) \equiv 2^{m-k-1}\pi^{-1/2}\Gamma(\frac{m-k}{2})\Gamma(\frac{m-k+1}{2})$ from equation (\ref{dup_gamma}), $f(t+1,m)$ is

\begin{align*}
f(t+1,m) &= [\frac{1}{\pi^{(t-1)/4}}
\frac{\Gamma(\frac{m}{2})}{\Gamma(\frac{m-t+1}{2})}] \times
1 \times
\prod_{k=1}^{t-1}[
\frac{1}{\pi^{-1/4}}
\frac{\Gamma(\frac{m-k}{2})}{\Gamma(\frac{m-k+1}{2})}
]    \\
&= \prod_{k=1}^{t}[
\frac{\Gamma(\frac{m-k+1}{2})}{\Gamma(\frac{m-k+1}{2})}
]  \nonumber \\
&= 1  \qedhere
\end{align*}
\end{proof}

\section{Computational time analysis}
Lewandowski et al. (\citeyear{Lewandowski2009}) proposed the onion method to generate random correlation matrices from LKJ distribution. The preceding section proposed another method to generate random matrices from an RW distribution with m degrees of freedom $RW_{T}(m)$. According to the Bartlett decomposition (\citealp{Smith1972}), if X follows the Wishart distribution with scale matrix $\boldsymbol{I}_{T}$ (T-dimensional Identity matrix) and m degrees of freedom, $\boldsymbol{X} \sim W_{T}(m,\boldsymbol{I}_{T})$, then $\boldsymbol{X}=\boldsymbol{A}\boldsymbol{A}^{\prime}$, where
\begin{align}
 \boldsymbol{A} = \begin{pmatrix}
	c_1 & 0 & 0 & \cdots & 0\\
	n_{21} & c_2 &0 & \cdots& 0 \\
	n_{31} & n_{32} & c_3 & \cdots & 0\\
	\vdots & \vdots & \vdots &\ddots & \vdots \\
	n_{T1} & n_{T2} & n_{T3} &\cdots & c_T
\end{pmatrix}
\end{align}
where $c^2_{i} \sim \chi^2_{m-i+1}$ and $n_{ij} \sim N(0,1)$ independently.
Then to generate random matrices from $RW_{T}(m)$, we first generate matrix $\boldsymbol{A}$, then calculate $\boldsymbol{X}=\boldsymbol{A} \boldsymbol{A}^{\prime}$, and finally get random correlation matrix $\boldsymbol{P}_{1}$ based on covariance matrix $\boldsymbol{X}$. Similarly, to generate random matrices from $RIW_{T}(m)$, we first generate matrix $\boldsymbol{A}$, then by solving a lower triangular system of linear equations we get $\boldsymbol{B}=\boldsymbol{A}^{-1}$, and finally get random correlation matrix $\boldsymbol{P}_{2}$ based on covariance matrix $\boldsymbol{Y}=\boldsymbol{B}^{\prime}\boldsymbol{B}$. Although the time complexity of the three methods (RW, RIW and onion method) are all constrained by LU (lower-triangular upper-triangular) matrix multiplication, to investigate the difference we compared them in R (\citealp{R2008}) on a server with a Haswell E5-2680v3 processor and 998 GB of RAM memory. The R code is in the supplementary file.

Table 1 list times (in seconds) needed to generate 5000 random correlation matrices of given dimension $T$ and degrees of freedom $m=T+1$. The difference between the onion method and the RW method is small when dimension is large ($T>240$). However, the RW method is much faster than the onion method for low dimension ($T<120$). The RW and RIW method have comparable speed.
\begin{table}
	\centering
	\begin{tabular}{l c c c}
		\hline
		Dimension & LKJ & RW & RIW \\
		\hline
		20 & 1.53 & 0.70 & 0.80 \\
		40 & 3.37 & 1.46 & 1.44 \\
		80 & 8.44 & 5.03 & 5.06 \\
		120 & 16.90 & 12.26 & 9.85 \\
		200 & 34.40 & 28.78 & 29.08 \\
		240 & 47.39 & 44.12 & 42.90 \\
		280 & 66.37 & 62.59 & 58.85 \\
		\hline
	\end{tabular}
	\caption{Time (in seconds) needed to generate 5000 correlation matrices of given dimension}
	\label{Computational Time}
\end{table}

\section{Conclusion}
This paper's main aim was to show the equivalence of the restricted Wishart distribution and the LKJ distribution. Such equivalence not only helps us understand why a Wishart prior might work for a correlation matrix but also indicates another way (partial correlation) to understand Wishart distribution. We also proposed another way to generate random correlation matrices, which is somewhat faster than the onion method by Lewandowski et al.
\section*{Acknowledgments}
Thanks to Jim Hodges for help with the English prose.
\bibliography{ref_zhenxun}
\bibliographystyle{kluwer}

\vfill
\eject

%\begin{figure}[t!]
%\centering
%\includegraphics[height=150mm]{salivaPlotV2.pdf}
%\caption{}
%\label{fig:disCI}
%\end{figure}

\end{document}